 \documentclass[twocolumn,showpacs,amsmath,amssymb]{revtex4-1}

\usepackage{graphicx}
\usepackage{dcolumn}
\usepackage{bm}
\usepackage{epsfig}
\usepackage{color}
\usepackage{longtable}
\usepackage{datetime}

\def\be{\begin{equation}}
\def\ee{\end{equation}}
\def\bea{\begin{eqnarray}}
\def\eea{\end{eqnarray}}
\def\ba{\begin{array}}
\def\ea{\end{array}}
\def\Term#1 #2 #3/{\mbox{$\,^{#1}\!#2_{#3}$}}
\def\Termo#1 #2 #3/{\mbox{$\,^{#1}\!#2^o_{#3}$}}

\begin{document}

\title{Toward calculations with spectroscopic accuracy: the $2s^22p$~$^2P^o_{3/2}$ -- $2s2p^2$~$^4P_{5/2}$ Excitation  Energy in Boron}

\author{Charlotte Froese Fischer}
\email{fischer@nist.gov}
\affiliation{National Institute of Standards and Technology,  Gaithersburg, MD 
20899-8422, USA}

\author{Simon Verdebout}
\author{Michel Godefroid}
\affiliation{Chimie Quantique et Photophysique, Universit\'e Libre de Bruxelles, B-1050 Brussels, Belgium}

\author{Pavel Rynkun}
\affiliation{Lithuanian University of Educational Science, Student\c u 39, LT-08106 Vilnius, Lithuania}

\author{Per J\"onsson}
\affiliation{School of Technology, Malm\"o University, S-205~06 Malm\"o, Sweden}

\author{Gediminas Gaigalas}
\affiliation{Institute of Theoretical Physics and Astronomy, Vilnius University, LT-01108 Vilnius, Lithuania}

\date{\today \; \; -  \currenttime \\[0.3cm]}

\begin{abstract}
No lines have been observed for transitions between the doublet and quartet levels of B~{\sc i}. 
Consequently, energy levels based on observation for the latter are obtained 
through extrapolation of wavelengths along the iso-electronic sequence for
 the $2s^22p$ \Termo 2 P {3/2}/ --  $2s2p^2$ \Term 4 P {5/2}/ transition.
In this paper, accurate theoretical excitation energies from a newly developed 
partitioned correlation function interaction (PCFI) method are reported for B~{\sc i}
that include both relativistic effects in the Breit-Pauli
approximation and a finite mass correction.  Results are compared with extrapolated values
from observed data. For B~{\sc i} our estimate of the excitation energy, 
28959$\pm$5~cm$^{-1}$, is in better agreement with the values obtained by Edl\'en {\it et al.} 
(1969) than those reported by Kramida and Ryabtsev (2007).  Our method is validated by applying 
the same procedure to the separation of these levels in C~{\sc ii}. 

\end{abstract}

\maketitle

\section{Introduction} \label{sec:intro}

Intercombination lines are not observed in B~{\sc i}, hence the position of the
quartets relative to the doublets is obtained by the extrapolation of wavelengths from
observed
data for the $2s^22p \Termo 2 P {3/2}/ - 2s2p^2$ \Term 4 P {5/2}/ transition
along the iso-electronic sequence.
Edl\'en {\it et al.}~\cite{edlen} relied on non-relativistic theory to 
scale available excitation energies by an effective nuclear charge $Z-2.97$, where
$Z$ is the atomic number. The screening parameter of 2.97 was chosen to yield a 
curve with minimum variation. Edl\'en {\it et al.} estimated 
the energy separation of the levels in B~{\sc i} to be 28866 $\pm$ 15 cm$^{-1}$.  
Recently, Kramida and Ryabtsev~\cite{KR} revised the position of the 
$2s2p^2$ \Term 4 P {5/2}/ level, using an extended set of experimental data
over the range $Z=6-14$.  Their revised estimate for the energy separation
 is 28643.1 cm$^{-1}$ with an 
uncertainty of 1.8 cm$^{-1}$ that does not include the extrapolation error.
This revised value is now included in the Atomic Spectra Database 
(ASD)~\cite{asd}.
The difference in the Edl\'en {\em et al.} and Kramida and Ryabtsev values 
is 223 cm$^{-1}$.

Multiconfiguration Hartree-Fock calculations with relativistic corrections in the
 Breit-Pauli (MCHF+BP) approximation have been reported 
for the boron-like iso-electronic sequence by Tachiev and 
Froese Fischer~\cite{TFF}. 
For B~{\sc i}, C {\sc ii}, and N~{\sc iii},  the computed \Term 4 P {}/ 
levels were too
 {\em high} relative to ASD values~\cite{asd} of that time by 
(0.53, 0.13, 0.10) \%, respectively.
 With the revision (now adopted in ASD), this error in  
B~{\sc i} 
has increased to 1.3~\% or 277~cm$^{-1}$. 
  Such a large error for boron 
compared with the error for  other elements is surprising.

Boron is a relatively light atom in which the fine-structure splitting is small
and the spin-orbit interaction between different
$LS$ terms is negligible. For this reason,
earlier ``spectrum'' calculations ( for energy levels relative to the ground state
rather than total energies) 
were often restricted to doublet terms~\cite{PJ+CFF}.  The MCHF+BP
calculation by Tachiev and Froese Fischer
 was intended as a spectrum calculation for all levels of the lower 
portion of the spectrum for the iso-electronic sequence, with the neutral atom
not the element of prime concern. 
In particular, correlation in the $1s^2$ core was omitted 
since its contribution to the wave function in the outer region 
of the atom decreases rapidly along the iso-electronic sequence.
With current computers and techniques for treating correlation,
 calculations of  considerably higher accuracy are feasible.  

For light atoms, Hylleraas-CI methods are often the most accurate, and
considerable progress has been made for the $^2\!P^o$ ground state 
energy~\cite{ruiz} of boron, but it is not yet the most accurate.
Using state of the art configuration interaction methods with energy optimized
Slater type orbitals,  Almora-D\'iaz and Bunge~\cite{Diaz}
obtained a non-relativistic
 ground state energy of $-24.653861(2)~E_{\rm h}$ (Hartree units), with an uncertainty of less
than 0.5~cm$^{-1}$. 
Recently results from 
correlated-Gaussian ~\cite{Bubin} 
 calculations have been reported with a slightly
lower energy of $-$24.65386608(250)~$E_{\rm h}$ for a nucleus of infinite mass,
 an energy that increased when the finite mass effect
 was included. 
The latter, very detailed calculation, is in a class of its own,
requiring one year 
of continuous computing on a multiprocessor system employing 16 -- 24 cores.  
In neither of these papers were results available 
for the $^4\!P$ excited state and relativistic effects were not considered. 
The ``exact'' non-relativistic total energy has been derived from observed 
data to be  $-$24.65393~$E_{\rm h}$~\cite{chakravorty}.

Relativistic energies relative to the ground state 
have been reported by Safronova {\it et al.}~\cite{SJS}
obtained by applying perturbation
theory to the calculation of energies for various values of $Z$. 
For boron, the energy of 
{$2s2p^2$ \Term 4  P {5/2}/ relative 
to  $2s^22p$ \Termo 2 P {3/2}/  } 
 was 29688~cm$^{-1}$, considerably higher
than the Kramida and Ryabtsev value.
{
This excitation energy can be calculated from Gu's total energies~\cite{Gu}  
as an even higher value of  29917.06 cm$^{-1}$ 
from a fully relativistic,
combined configuration interaction and many-body perturbation theory 
method, but does not claim high accuracy.

 In this paper we report the results of calculations
for the
$2s^22p$ \Termo 2 P {3/2}/ - $2s2p^2$ \Term 4 P {5/2}/ excitation energy, 
both in the 
MCHF+BP method tailored to the  neutral atom, and results from a newly 
developed partitioned correlation function interaction (PCFI) method that allows
non-orthogonal orbital bases for the inclusion of  different 
correlation effects~\cite{Veretal:2010a,Veretal:2013a}. Relativistic 
 and finite mass corrections are included. Accuracy 
is estimated and validated by similar calculations for C~{\sc ii} where
experimental data are available.

\section{The MCHF method: theory and results}

In the multiconfiguration Hartree-Fock method~\cite{book_atsp}, the wave function 
$\Psi(\gamma LS)$ for the state labeled  $\gamma LS$ is a linear combination
of configuration state functions (CSFs) defined in terms of an orthonormal
orbital basis. The {\sc atsp2k} code~\cite{atsp2k} can be used to determine the optimal basis
and CSF expansion coefficients that define a stationary energy.  In the present
case, since both states are lowest in their symmetry,  the optimal solution
 minimizes the total energy of the state.  The accuracy of the computed
energy relative to the exact energy of the state depends largely on the CSFs 
included in the expansion and the orbital basis.

In this work, systematic calculations were performed in which the orbital basis 
was increased from one calculation to the next by increasing the maximum principal 
quantum number $n$, thereby introducing a new ``layer'' of
orbitals. This parameter $n$ 
characterizes the calculation. MCHF calculations up to $n=9$ were
performed with orbital quantum numbers up to $l=5$ ($h$-orbitals).

The computational model determines the CSF space for each orbital set.
The {\sc lsgen} program~\cite{lsgen} was used to generate all CSFs that differ 
by one or more electrons
from a given configuration and are of the same parity and $LS$ quantum number.
Such substitutions are referred to as single- (S), double- (D), etc. excitations.
For open-shell systems, SD excitations from a single configuration are not 
sufficient, as shown for the Be ground state~\cite{Veretal:2010a,book_atsp}.
On the other hand, a Complete Active Space (CAS) expansion for which the number of 
excitations equals the number of electrons, soon becomes impractical
as the orbital set increases. Here we
have used a method, referred to as MR-SD,  that starts with a multireference set 
(MR) of CSFs that 
includes the CSF for the state and any others that may be important.  Then SD
expansions are obtained by applying the process to each member of the MR set.

Ideally, the MR set should contain all the important CSFs of the final wave function.
An indication of these CSFs can be obtained from a small MCHF calculation, such as
an MCHF calculation for an $n=3$ SD expansion.  Those with small expansion coefficients
can be omitted, but as more correlation is added, others may gain importance.
In the present work, the default MR sets are:
\begin{eqnarray}
\label{eqn:MR_SD}
^2\!P^o&:& 2s^22p, 2p^3, 2s2p3d, 2s^23p, 2s3s3p, 2p^23p, 2p3s^2,\nonumber \\ 
       & & 2p3d^2, 2s2p3s  \\
^4\!P  &:& 2s2p^2, 2s2p3p, 2p^23d, 2s3d^2, 2s3p^2, 2p3s3p, 2p^23s \nonumber
\end{eqnarray}
\noindent
In the final wave function, the smallest expansion coefficient was about 0.025 in magnitude, 
accounting for $\approx 0.06$ \% of the {eigenvector} composition.

  
The present two states of interest have a common $1s^2$ core and three valence
electrons.  The substitution of one or two valence orbitals
by other orbitals defines valence-valence (VV) correlation, 
substitution of a single $1s$ orbital and possibly also one valence orbital 
defines core-valence (CV) correlation, and 
substitution of the two $1s$ orbitals defines core-core (CC) correlation. 
 Correlation in the core essentially
cancels in the calculation of an energy difference, but it may be large. Therefore 
small differences may contribute significantly to the energy separation.  In 
order to maintain this balance at intermediate stages, prior to convergence,
the variational principle was applied to the sum of
energy functionals for each state, referred to as simultaneous optimization.  
Since both states are lowest in their symmetry, the sum of the energies 
is also a minimum.  In this way, the same orbitals were used
in the calculation of the core of both states and the optimization process 
minimized the combined energy.


\begin{table}
\caption{Comparison of the $2s^22p$ \Termo 2 P {}/ - $2s2p^2$ \Term 4 P {}/ 
excitation energy, $\Delta E$ in 
 cm$^{-1}$, for various computational strategies (see text for details). The number of
CSFs, $N$ and the total 
energies $E$ (in $E_{\rm h}$) {corresponding to the largest active set ($9h$) wave functions}
for the \Termo 2 P {}/ and \Term 4 P {}/ terms are 
given at the end of the table.\\   
}
\label{results}
\begin{tabular}{@{} l  c c c c c  @{}}
\hline \hline \\ [0.02cm]
$n$ &  $\Delta E$([A])  &  $\Delta E$([B]) & $\Delta E$([C]) & $\Delta E$([D]) & $\Delta E$([E]) \\
\hline \\[0.01cm]
4     &   28301.91 & 28318.01 & 28319.91 & 28382.21 & 28677.73 \\
5     &   28424.82 & 28428.62 & 28431.02 & 28755.73 & 28658.93 \\
6     &   28672.23 & 28689.63 & 28691.43 & 28892.64 & 28768.53 \\
$7h$  &   28762.13 & 28780.53 & 28781.73 & 28945.14 & 28816.54 \\
$8h$  &   28819.34 & 28824.04 & 28824.94 & 28967.24 & 28835.64 \\
$9h$  &   28846.64 & 28851.74 & 28852.84 & 28978.64 & 28845.74 \\
[0.1cm]
\hline 
\Termo 2 P {}/ \\
$N$ & 74,103    & 74,103 & 158,337 & 35,919 & 35,919 \\
$E$  & $-$24.65289 & $-$24.65286 & $-$24.65290 & $-$24.61069 & $-$24.62290 \\
\Term 4 P {}/ \\
$N$ & 50,095 & 50,095 & 113,935 & 25,250 & 25,250 \\
$E$  & $-$24.52145 & $-$24.52141 & $-$24.52144 & $-$24.47856 & $-$24.49147\\
[0.1cm]
\hline\hline \\
\end{tabular}
\end{table}

Table~\ref{results} shows the variation of the 
\Termo 2 P {}/ - \Term 4 P {}/ excitation energy for several different
computational strategies all using the MCHF method in which all orbitals are
optimized unless specified to the contrary.

 In the {\sl independent} strategy (used in calculation [A]), 
the wave function for each state is calculated independently 
in a single orthonormal orbital basis that then describes all three types of correlation. 
As shown previously~\cite{Veretal:2010a}, the orbitals for
VV correlation will have a maximum in the outer region of the atom,
 the CV orbitals
are in the regions of overlap between $1s$ and $2s$ or $2p$, and CC is in the region of the 
$1s$ orbital. At each stage of a systematic calculation, the orbitals arrange themselves
so as to minimize the energy and, in going from one layer to the next, considerable 
rearrangements may occur. 
If the rearrangements of the orbitals in the 
two independent calculations are not similar at an intermediate stage, an imbalance {\it may} occur in the difference. 

In the {\sl simultaneous}  strategy (used in calculation [B]), the variational
 procedure is applied to the 
sum of the two energies so that the same orbitals are used to describe both
 wave functions. 
In particular, the more dynamic CC correlation is described in terms of the 
same orbitals.
 Table~\ref{results} shows that, with the same MR-SD expansion, the simultaneous optimization 
method closely tracks calculation [A], although the excitation energy is somewhat larger. 
No large differences were found, even though the total energy of the \Term 4 P {}/ was raised 
more than that of \Termo 2 P {}/, thereby increasing $\Delta E$, but the difference 
decreased with~$n$. 
In order to test the dependence of the final 
answers on the MR set, calculation [C]  repeated the simultaneous optimization,
 starting with an $n=3$ MR set that included SDTQ excitations from all shells. 
Though the
expansions are much larger, the final excitation energy is only 1.10 cm$^{-1}$  larger
and the total energies less than 4.0$\times 10^{-5}$~$E_{\rm h}$ lower than calculation [B].  
The radii for all the $9l$  optimized orbitals
of the $n=9h$ layer calculation were CC orbitals as well as some of the $8l$ orbitals,
showing orbital dependence on all three types of correlation. 

  In calculation [D], the CSFs with an unoccupied $1s$-shell were omitted from the expansion,
unless the remaining shells consisted only of $2s,2p$ electrons, 
thereby neglecting CC correlation outside the $n=2$ complex. 
Simultaneous optimization was applied.  The occupied orbitals, 
($1s, 2s, 2p$) were obtained from a small MCHF calculation and then kept fixed.
The final $n=9h$ result is similar to the Tachiev and Froese Fischer~\cite{TFF}
value and because the correlation that is included is over a smaller region of space, 
the convergence with respect to the orbital set is rapid.  The fully variational results with 
the same expansion, and simultaneous
optimization (calculation [E]),  had the most convergence
problems of any method. A considerable transformation of the orbitals occurred,
and it was important 
not to require  that the $1s, 2s, 2p$ orbitals be spectroscopic orbitals with 
the usual nodal structure.  In particular, the final $1s$ orbital was 
contracted and, in the screened hydrogenic 
model, had a screening parameter of $-$0.14. 
The final excitation energy, however, was in excellent agreement with the
independently optimization results.  

Table~\ref{results} also specifies the size ($N$)  of the {($9h$)} wave function expansions
for each calculation.  Those without core correlation ([D] and [E])
have much smaller wave function expansions, a fact this is important in 
atoms with a large core. The total energy of each strategy is reported as 
well. Omitting CC raises the total energy significantly but the effect on
the excitation energy is much smaller. This table clearly indicates that the correct excitation
energy is closer to the Edl\'en {\sl et al.} value than the Kramida and Ryabtsev one.

Extending the MCHF method to higher accuracy would require a rather large set 
of orthonormal orbitals {since orbitals with} higher angular quantum 
number{s} should also be included.
An alternative is to introduce the use of non-orthogonal orbitals. 
\section{The Partitioned Correlation Function Interaction}

The partitioned correlation function interaction (PCFI) method that 
has been described in detail elsewhere~\cite{Veretal:2010a,Veretal:2013a}, 
 differs from the MCHF method in that the correction to the wave function,
$\Psi_0$, of the MR space is partitioned so that different orbital bases can be used for 
different correlation effects.

In this method the correction to the wave function for three types of
correlation is a linear combination of partitioned correlation functions (PCF's), denoted
by $P_i$, so that
\be
\Psi_1 = c_1P_1 +c_2 P_2 + c_3P_3 .
\ee

Each $P_i$ is itself a linear combination of CSF's, where the expansion coefficients and 
the orbitals are an  MCHF solution for the wave function $\Psi_0 + P_i$, 
with the orbitals of $\Psi_0$ fixed. 
Because the MCHF method requires a single orthonormal basis, the new orbitals for each 
partition $P_i$ must be orthogonal to the orbitals defining $\Psi_0$, but
may be non-orthogonal to those from other partitions. For the
 present case, the partitions represent VV, CV, and CC correlation corrections, 
respectively.
Biorthogonal methods are then used to compute the interaction matrix in a  non-orthogonal
basis for a wave function expanded in the CSF basis 
defining $\Psi_0$, and the partitions, $P_1, P_2, P_3$. Some variational freedom is lost by
this method since the expansion coefficients for each partition
 can now only be scaled by a constant factor.
In order to recover this freedom, a de-constraining procedure may be
 applied in which, when forming the PCFI configuration interaction 
matrix, a configuration state function is moved from a partition into the set defining 
$\Psi_0$~\cite{Veretal:2013a}.

\subsection{Non-relativistic Calculations for Boron}
The same MR sets as used for the MCHF study were adopted.
Applying the SD process to each set, the resulting expansions were classified 
according to the occupation number of the $1s$-shell. For each state, the next step was to obtain 
the orbitals for $\Psi_0$.  Keeping these orbitals frozen,
MCHF calculations were performed for each of their three 
partitions.  Because the orbital basis now targets the same type of correlation, 
there is less rearrangement of orbitals as new layers of orbitals are added,
 so that convergence is better and 
calculations can readily be extended to $n=10$ without any truncation of the 
angular quantum numbers.
 
Using the resulting PCFs, the configuration interaction matrix was built.
Table~\ref{tab:B_MR1_cont} displays the total energies and excitation energy
 calculated with the original constrained representation of each PCF.
The energies of the $n=9$ results are significantly lower than the MCHF results
of Table~\ref{results}.
 
To check the sensitivity of the excitation energy to the constraint effect, 
we also report results from the 
De-constrained Partitioned Correlation Functions Interaction (DPCFI) method 
where all the expansion coefficients are free to vary.
The modified excitation energy is included in Table~\ref{tab:B_MR1_cont} 
in the last column and denoted as $\Delta E$(D). The final 
total energies and excitation energy are presented as $n=10D$ results.
From these data we see that the de-constraint increases the excitation energy by
6.39~cm$^{-1}$.  
The changes are small but bring the total energy into better agreement with 
the ``exact'' non-relativistic energy~\cite{chakravorty} for the ground state.
%

\begin{table} [!ht]
\caption{\label{tab:B_MR1_cont} Non-relativistic total energies $E$ (in $E_{\mathrm{h}}$) 
and excitation energy, $\Delta E$ in cm$^{-1}$, of the lowest $^2P^o$ and $^4P$ terms of neutral 
boron obtained with the PCFI and DPCFI method. 
 The total energies for a DPCFI calculation are reported
as 10D and the excitation energy as $\Delta E(D)$.  }
\begin{center}
\begin{tabular}{c  r r c c}\\
\hline \hline \\[-0.2cm]
$n$ & \multicolumn{1}{c}{$E$ $^2P^o$} &\multicolumn{1}{c}{$E$ $^4P$} & \multicolumn{1}{c}{$\Delta E$} &
\multicolumn{1}{c}{$\Delta E$(D)} \\[0.1cm]
\hline \\[-0.2cm]
4 	 & $-24.644046909$ 	 & $-24.514398526$ 	 & $28454.53$ & 28457.66 \\
5 	 & $-24.650782918$ 	 & $-24.519740851$ 	 & $28760.40$ & 28765.41\\
6 	 & $-24.652391314$ 	 & $-24.520979432$ 	 & $28841.57$ & 28847.00\\
7 	 & $-24.652978136$ 	 & $-24.521441671$ 	 & $28868.91$ & 28875.86 \\
8 	 & $-24.653244882$ 	 & $-24.521636963$ 	 & $28884.59$ & 28890.74\\
9 	 & $-24.653383550$ 	 & $-24.521735733$ 	 & $28893.35$ & 28899.63\\
10 	 & $-24.653464335$ 	 & $-24.521792201$ 	 & $28898.69$ & 28905.08\\[0.1cm]
10D      & $-24.653523595$    & $-24.521822334$    &      & $28905.08$\\[0.1cm]
\hline \\[-0.2cm]
$E^{\hbox{\tiny{exact}}}$\cite{chakravorty} & $-$24.65393 \\
\hline \hline
\end{tabular}
\end{center}
\end{table}

The effect of increasing the MR set was also evaluated, using the 
complete active space (CAS) concept for a valence correlation
 expansion  of $\Psi_0$ for each of the two states, namely
\begin{eqnarray}
\label{MR_CAS}
1s^22s^22p~^2P^o : \qquad  1s^2\{2s,2p,3s,3p,3d\}^3~^2P^o \nonumber\\
1s^22s2p^2~^4P   : \qquad  1s^2\{2s,2p,3s,3p,3d\}^3~^4P\;. 
\end{eqnarray}
This notation represents an expansion over CSFs from 
configurations with a $1s^2$ core and  
three orbitals of the required symmetry and parity for the given list of 
orbitals. The MR expansions
contain 30 and 13 CSFs for the odd and even parity, respectively. 
Allowing single and double excitations from these CSFs we get the 
CSF space that was partitioned into the three PCFs (VV, CV, and CC).
The expansion size
grows rapidly -- 242,532 and 175,542 for \Termo 2 P {}/ and \Term 4 P {}/,
respectively for $n=9$ and 357,230 and 258,565 for $n=10$.
However, with partitioning the expansion into three parts, 
the final size is computationally feasible. 

Systematic calculations can be improved through extrapolation of 
trends to $n=\infty$ which in our procedure also implies $l=\infty$.
MCHF convergence trends have been investigated by Tong {\it et al.}~\cite{tong}
 where results are extrapolated
first by $n$, as a function on the maximum $l$, and then for $l$. The rate
of convergence of the former for a given maximum $l$ is more rapid than 
for ${l}$.
For helium, Schwartz~\cite{schwartz} showed that the contribution 
from higher
orbital angular momenta was proportional to $1/l^4$ for symmetric states 
and $1/l^6$ for antisymmetric. For this angular type of asymptotic behavior, the rate
of convergence $(E_{l+1}-E_l)/(E_{l}-E_{l-1})$ approaches unity. 

In the present work, a much simpler procedure has been used to estimate the 
remaining uncalculated contribution, a procedure that underestimates the correction.
To perform the extrapolation, we consider the change in energy (or 
excitation energy)  $\delta E_9 = E_9 - E_8$ and $\delta E_{10} = E_{10}- 
 E_9 $
 to determine the rate of convergence :
\begin{equation}\label{eq:ext1}
r = {\delta E_{10}}/{\delta E_9}\;.
\end{equation}
If we assume that the convergence continues at the same rate for the rest of the
sequence, the remainder is a geometric series that can be summed to yield
\begin{equation}\label{eq:ext2}
\sum_{i=1}^{\infty} \delta E_{10}\, r^{i} = \delta E_{10}\, \frac{r}{1-r} \hspace*{1cm}\textrm{ if } |r| < 1\;,
\end{equation}
In fact, the ratio increases with $n$ because of the slow convergence of contributions from 
the higher angular momenta.

\begin{table} [!h]
\caption{\label{tab:BMR2_contRS} PCFI total energies, $E$ in $E_{\mathrm{h}}$, and 
excitation  energy, $\Delta E$ in cm$^{-1}$ including the relativistic shift operator, of the lowest 
$^2P^o$ and $^4P$ terms of neutral boron. The MR set
included all CSFs of $1s^2\{2s,2p,3s,3p,3d\}^3$ of the required symmetry and parity. $\Delta E$(S) refers to the excitation energy including the relativistic 
shift operator from calculations with the smaller MR of Table II. $N$ is the size of the $n=10$ CSF
expansion.}
\begin{center}
\begin{tabular}{c r r c c }\\
\hline \hline \\[-0.2cm]
$n$ & \multicolumn{1}{c}{$E$ $^2P^o$} &\multicolumn{1}{c}{$E$ $^4P$} 
& \multicolumn{1}{c}{$\Delta E$ } 
& \multicolumn{1}{c}{$\Delta E$(S) }  \\[0.1cm]
 \hline \\[-0.2cm]
4 	& $-24.650382640$ 	& $-24.520438740$ & $28519.38$ & $28496.83$\\
5 	& $-24.657117561$ 	& $-24.525847143$ & $28810.52$  & $28803.21$\\                                                                                                                                                                                                                                                                                                                                                                                                                                                                                                                                                                                                                                                                                                                                                                                                                                                                                                                                                                                                                                                                                                                                                                                                                                                                                                                                                                                                                                                                                                                                                                                                                                                                                                                                                                                                                                                                                                                                                                                                                                                                                                                                                                                                                                                                                                                                                                                                                                                                                                                                                                                                                                                                                                                                                                                                                                                                                                                                                                                                                                                                                                                                                                                                                                                                                                                                                                                                                                                                                                                                                                                                                                                                                                                                                                                                                                                                                                                                                                                                                                                                                                                                                                                                                                                                                                                                                                                                                                                                                                                                                                                                                                                                                                                                                                                                                                                                                                                                                                                                                                                                                                                                                                                                                                                                                                                                                                                                                                                                                                                                                                                                                                                                                                                                                                                                                                                                                                                                                                                                                                                                                                                                                                                                                                                                                                                                                                                                                                                                                                                                                                                                                                                                                                                                                                                                                                                                                                                                                                                                                                                                                                                                                                                                                                                                                                                                                                                                                                                                                                                                                                                                                                                                                                                                                                                                                                                                                                                                                                                                                                                                                                                                                                                                                                                                                                                                                                                                                                                                                                                                                                                                                                                                                                                                                                                                                                                                                                                                                                                                                                                                                                                                                                                                                                                                                                                                                                                                                                                                                                                                                                                                                                                                                                                                                                                                                                                                                                                                                                                                                                                                                                                                                                                                                                                                                                                                                                                                                                                                                                                                                                                                                                                                                                                                                                                                                                                                                                                                                                                                                                                                                                                                                                                                                                                                                                                                                                                                                                                                                                                                                                                                                                                                                                                                                                                                                                                                                                                                                                                                                                                                                                                                                                                                                                                                                                                                                                                                                                                                                                                                                                                                                                                                                                                                                                                                                                                                                                                                                                                                                                                                                                                                                                                                                                                                                                                                                                                                                                                                                                                                                                                                                                                                                                                                                                                                                                                                                                                                                                                                                                                                                                                                                                                                                                                                                                                                                                                                                                                                                                                                                                                                                                                                                                                                                                                                                                                                                                                                                                                                                                                                                                                                                                                                                                                                                                                                                                                                                                                                                                                                                                                                                                                                                                                                                                                                                                                                                                                                                                                                                                                                                                                                                                                                                                                                                                                                                                                                                                                                                                                                                                                                                                                                                                                                                                                                                                                                                                                                                                                                                                                                                                                                                                                                                                                                                                                                                                                                                                                                                                                                                                                                                                                                                                                                                                                                                                                                                                                                                                                                                                                                                                                                                                                                                                                                                                                                                                                                                                                                                                                                                                                                                                                                                                                                                                                                                                                                                                                                                                                                                                                                                                                                                                                                                                                                                                                                                                                                                                                                                                                                                                                                                                                                                                                                                                                                                                                                                                                                                                                                                                                                                                                                                                                                                                                                                                                                                                                                                                                                                                                                                                                                                                                                                                                                                                                                                                                                                                                                                                                                                                                                                                                                                                                                                                                                                                                                                                                                                                                                                                                                                                                                                                                                                                                                                                                                                                                                                                                                                                                                                                                                                                                                                                                                                                                                                                                                                                                                                                                                                                                                                                                                                                                                                                                                                                                                                                                                                                                                                                                                                                                                                                                                                                                                                                                                                                                                                                                                                                                                                                                                                                                                                                                                                                                                                                                                                                                                                                                                                                                                                                                                                                                                                                                                                                                                                                                                                                                                                                                                                                                                                                                                                                                                                                                                                                                                                                                                                                                                                                                                                                                                                                                                                                                                                                                                                                                                                                                                                                                                                                                                                                                                                                                                                                                                                                                                                                                                                                                                                                                                                                                                                                                                                                                                                                                                                                                                                                                                                                                                                                                                                                                                                                                                                                                                                                                                                                                                                                                                                                                                                                                                                                                                                                                                                                                                                                                                                                                                                                                                                                                                                                                                                                                                                                                                                                                                                                                                                                                                                                                                                                                                                                                                                                                                                                                                                                                                                                                                                                                                                                                                                                                                                                                                                                                                                                                                                                                                                                                                                                                                                                                                                                                                                                                                                                                                                                                                                                                                                                                                                                                                                                                                                                                                                                                                                                                                                                                                                                                                                                                                   
6 	& $-24.658725942$ 	& $-24.527084872$ & $28891.87$ & $28885.02$\\
7 	& $-24.659297417$ 	& $-24.527551340$ & $28914.92$ & $28912.18$\\
8 	& $-24.659568315$ 	& $-24.527751544$ & $28930.43$ & $28927.90$ \\
9 	& $-24.659709842$ 	& $-24.527853807$ & $28939.05$ & $28936.71$\\
10 	& $-24.659792829$ 	& $-24.527912772$ & $28944.32$ & $28942.06$ \\[0.1cm]
$\infty$ & $-24.659910$ & $-24.527993$ & $28952.52$ \\
\hline
$N$ &  357,230 & 258,565 \\
\hline \hline
\end{tabular}
\end{center}
\end{table}


\subsection{Relativistic and Finite Mass  corrections}

For a light atom, relativistic effects can  be accurately estimated in the 
Breit-Pauli approximation.
The terms of the  Breit-Pauli operator
 can be classified into the $J$-dependent
fine-structure (FS) and $LS$-dependent relativistic shift (RS) 
contributions~\cite{book_atsp}.
For boron, the latter are the more important 
corrections and {were} easily included in this extensive calculation.

Table~\ref{tab:BMR2_contRS} presents the total energies of each state as 
well as the excitation energies obtained using the PCFI method with a
Hamiltonian including the relativistic shift operators. For the purpose of
comparison, excitation energies from the smaller calculations of 
Table~\ref{tab:B_MR1_cont} that
also include the relativistic shift operators are reported in the last column and 
are denoted as $\Delta E$(S). The increase in the excitation energy of 2.26 cm$^{-1}$ is in 
agreement with the small increase {observed in the MCHF calculations reported in 
Table~\ref{results} when including higher-order excitations ([B]$\rightarrow$[C]).} 
The RS contribution to the doublet-quartet excitation energy is estimated to 
be the difference between the $n=10$ values of
 28942.06 ($\Delta E(S)$ of Table~\ref{tab:BMR2_contRS}) and  28898.69 ($\Delta E$ of 
Table~\ref{tab:B_MR1_cont}),
or $43.37$~cm$^{-1}$.

Table~\ref{tab:B_MR1BPcont} reports the excitation energy using the complete 
Breit-Pauli Hamiltonian within the PCFI approach.  $LS$-term mixing  of
different terms is omitted.
{ 
Substracting the $\Delta E$(S) values reported in Table~\ref{tab:BMR2_contRS} from the Breit-Pauli $\Delta E$ values of Table~\ref{tab:B_MR1BPcont},
}
we get an estimate of the importance of the $LS$ diagonal fine structure operators, i.e. $-$0.74 cm$^{-1}$. 
Comparing the $\Delta E$ and $\Delta E(D)$ values in Table~\ref{tab:B_MR1BPcont} reveals the constraint effect on the excitation energy in the BP approximation, i.e. 7.05~cm$^{-1}$, which differs somewhat from the earlier estimate
of 6.39~cm$^{-1}$. The difference of 0.66~cm$^{-1}$ is small and establishes a lower limit on the uncertainty of our computational procedure.

\begin{table} [!h]
\caption{\label{tab:B_MR1BPcont}Breit-Pauli total energies, $E$ in $E_{\mathrm{h}}$, and 
corresponding excitation energy, $\Delta E$ in cm$^{-1}$, of the lowest $^2P_{3/2}^o$ and 
$^4P_{5/2}$ levels of neutral boron using both the PCFI and DPCFI methods.}
\begin{center}
\begin{tabular}{c l l c c c}\\
\hline \hline \\[-0.2cm]
$n$ & \multicolumn{1}{c}{$E(^2P_{3/2}^o)$} &\multicolumn{1}{c}
{$E(^4P_{5/2})$} &$\phantom{A}$& \multicolumn{1}{c}{$\Delta E$} 
& \multicolumn{1}{c}{$\Delta E$(D)} \\[0.1cm]
\hline \\[-0.2cm]
4 	 & $-24.650251936$ 	 & $-24.520414146$ 	 && $28496.10$ & $28499.14$\\
5 	 & $-24.657033103$ 	 & $-24.525799428$ 	 && $28802.46$ & $28807.83$\\
6 	 & $-24.658656372$ 	 & $-24.527049905$ 	 && $28884.28$ & $28890.09$\\
7 	 & $-24.659251843$ 	 & $-24.527521599$ 	 && $28911.44$ & $28918.01$\\
8 	 & $-24.659523392$ 	 & $-24.527721520$ 	 && $28927.16$ & $28933.87$\\
9 	 & $-24.659666371$ 	 & $-24.527824368$ 	 && $28935.97$ & $28942.94$\\
10 	 & $-24.659749687$ 	 & $-24.527883327$ 	 && $28941.32$ & $28948.37$\\[0.1cm]
$\infty$ & & & &  28949.56 & 28956.47 \\ [0.1cm]
10D 	 & $-24.659913687$ 	 & $-24.528015185$ 	 && &  $28948.37$\\
\hline \hline
\end{tabular}
\end{center}
\end{table}

The finite-mass correction {should also be considered}.  The normal mass (NMS) correction
can readily be determined from the Bohr mass scaling law, using
the finite-mass Rydberg constant. This reduces the excitation energy by $-1.59$ cm$^{-1}$ and 
$-1.44$ cm$^{-1}$
for $^{10}$B and $^{11}$B, respectively.
However, for the excitation energy under 
consideration the specific mass shift is larger than expected.  Using the 
$n=10$ PCFI wave functions for estimating the $\Delta S_{sms}$ difference of the specific mass shift parameters~\cite{Godetal:2001a}, the finite mass (NMS+SMS) corrections are:\\
\indent $^{10}$B:\quad  $-$6.67 cm$^{-1}$\quad and \quad $^{11}$B:\quad  $-$6.07 cm$^{-1}$\ \ . \\
{The de-constraint correction on $\Delta S_{sms}$ is very small and an average over the two isotopes based on the natural isotopic composition (19.9\% $^{10}$B/80.1\% $^{11}$B) gives a final estimation of $-$6.20 cm$^{-1}$. }
This correction is therefore important for spectroscopic accuracy.

\subsection{The C {\sc ii} quartet-doublet energy separation} 
\begin{table} [h]
\caption{\label{tab:CIIMR2_contRS}Total energies, $E$ in $E_{\mathrm{h}}$, and 
excitation energy,  $\Delta E$ in cm$^{-1}$ including the relativistic shift operator,
 of the lowest $^2P^o$ and $^4P$ terms of the singly ionized carbon atom obtained using the PCFI 
method as well as some DPCFI excitation energies.
  Observed data have been obtained from an $LS$ spectrum (see text). }
\begin{center}
\begin{tabular}{c c c c c}\\
\hline \hline \\[-0.2cm]
$n$ & \multicolumn{1}{c}{$E(^2P^o)$} &\multicolumn{1}{c}{$E(^4P)$} & 
\multicolumn{1}{c}{$\Delta E$} & \multicolumn{1}{c}{$\Delta E(D)$}  \\[0.1cm]
\hline\\[-0.2cm]
4 	& $-37.434442461$ 	& $-37.241016132$ & $42452.17$ & 42454.33 \\
5 	& $-37.441661608$ 	& $-37.246648961$ & $42800.32$ & 42804.62 \\
6	& $-37.443459176$ 	& $-37.247990596$ & $42900.39$ & 42905.57\\
7 	& $-37.444132978$ 	& $-37.248491723$ & $42938.29$ & 42944.35\\
8 	& $-37.444453444$ 	& $-37.248717425$ & $42959.09$\\
9 	& $-37.444620358$ 	& $-37.248830141$ & $42970.98$\\
10 	& $-37.444721702$ 	& $-37.248898421$ & $42978.24$\\[0.1cm]
$\infty$ 	
    & $-37.444878$ 	    & $-37.249003$ 	  & $42989.59$ \\
\multicolumn{3}{l}{De-constraint}                  &  \ \ 7.05 \\
\multicolumn{3}{l}{Finite Mass}                   & \ $-$10.22\\
\multicolumn{3}{l}{Excitation Energy}             & 42986.42 \\[0.1cm]
\hline\\[-0.2cm]
\multicolumn{2}{l}{Young {\it et al.}~\cite{Young:2011fk}} & $$ & \multicolumn{2}{l}{ $42993.0 \pm 0.9$}\\
ASD~\cite{asd} & $$ & $$ & \multicolumn{2}{l}{$42993.5$}\\[0.1cm]
\hline\hline
\end{tabular}
\end{center}
\end{table}


In order to estimate the errors not accounted for, mainly the contributions from 
orbitals with high-angular quantum numbers, we validate our method
by applying it to the calculation
of the excitation energy in C~{\sc ii} where the wavelength of the
\Termo 2 P {3/2}/ - \Term 4 P {5/2}/ transition has been measured
recently by Young {\it et al.}~\cite{Young:2011fk} and ASD~\cite{asd} values
are available. 

For maximum accuracy we start with the valence CAS expansion to determine 
$\Psi_0$ and
then the PCFs for the three types of correlation. In generating the 
configuration interaction matrix we used the Breit-Pauli Hamiltonian but with
only the relativistic shift operators, because of the size of the expansion
and the small effect from the $J$-dependent terms.  


Table~\ref{tab:CIIMR2_contRS} reports the results for C~{\sc ii}. Because of 
the large expansions, de-constrained DPCFI excitation energies are not included 
for the higher layers, but the effects of de-constraining closely track those of 
Table~\ref{tab:B_MR1BPcont} where the final difference was 7.05 cm$^{-1}$.
The finite mass correction {for an isotopically-unresolved line profile, 
is largely dominated by the lighest isotope (98.93\% $^{12}$C/1.07\% $^{13}$C) and
 is estimated to be $-10.22$~cm$^{-1}$. }
Correcting the extrapolated $n=10$
results  in Table~\ref{tab:CIIMR2_contRS} by these amounts
we get 42986.42 cm$^{-1}$.
Comparing this value with excitation energies derived from observed data
by defining each term energy to be the statistically weighted average of the
levels of the term,
 we get a remainder of {6.58}~cm$^{-1}$
representing residual correlation and other omitted effects that were not captured in our calculation for C~{\sc ii}.

\subsection{Final estimate for the $2s^22p$ 
\Termo 2 P {3/2}/ - $2s2p^2$ \Term 4 P {5/2}/ excitation energy in B~{\sc i}}

\begin{table}[h]
\caption{\label{final} Summary of contributions to the
$2s^22p$ \Termo 2 P {3/2}/ - $2s2p^2$ \Term 4 P {5/2}/ excitation energy
 (in cm$^{-1}$) in B~{\sc i} }
\begin{center}
\begin{tabular}{ l r l}\\
\hline \hline \\[-0.2cm]
Excitation energy (PCFI(MR)) & 28898.69\\
Relativistic shift (RS) & 43.37 \\
Fine-structure (FS)     & $-$0.74 \\
De-constraint   (D)      & 7.05 \\
Extrapolation  (X)		& 8.10 \\
Larger MR set (CAS)	& 2.26\\
Finite Mass 		& {$-$6.20} \\
Remainder (same as for C~\sc{ii})		& {6.58} \\
\hline  \\[-0.2cm]
Total Excitation Energy &  28959  & $\pm$ 5 \\
\hline  \\[-0.2cm]
{Edl\'en {\it et al.}~\cite{edlen} } & 28866 & $\pm$ 15 \\
{Kramida and Ryabstev~\cite{KR} } &   28643.1  & $\pm$ 1.8 \\
\hline \hline
\end{tabular}
\end{center}
\end{table}

The results of our investigation of the various aspects of the PCFI method as
applied to the $2s^22p$ \Termo 2 P {3/2}/ - $2s2p^2$ \Term 4 P {5/2}/ excitation energy are 
summarized in Table~\ref{final}. Listed are the contributions to 
the energy starting with the non-relativistic PCFI value of Table II based on the small MR list.
Since the corrections are small, values of contributions were obtained by expressing the 
$n=10$ excitation energies of
Tables~\ref{tab:B_MR1_cont}~-~\ref{tab:B_MR1BPcont} in terms of the PCFI value and 
contributions, assuming a first-order theory. As mentioned earlier,
 in some instances, a contribution
(such as D) could have two slightly different values, 
in which case it was determined from 
a calculation that included the most corrections in a given calculation.
  In fact, the sum of the 
first five entries is the extrapolated value of Table~\ref{tab:B_MR1BPcont} 
that includes the de-constraining correction.
 Our method of computing the correlation energy has accounted 
for the terms linear in $Z$ of a $Z$-dependent calculation.  Hence, remaining
correlation in B~{\sc i} should be similar to that in C~{\sc ii}. 
The uncertainty estimate largely represents the uncertainty of our 
estimate of the remainder.

 Thus our prediction for the 
$2s^22p$ \Termo 2 P {3/2}/ - $2s2p^2$ \Term 4 P {5/2}/ excitation energy is 
28959 cm$^{-1}$ $\pm$~5~cm$^{-1}$, considerably larger than the Edl\'en {\sl et al.}
value of 28866 $\pm$~15~cm$^{-1}$ and the Kramida and Ryabstev value of 28643.1 cm$^{-1}$.

\section{Conclusion}

The MCHF and PCFI methods have been combined to extend the accuracy of variational methods for 
complex atoms and ions. In this case, near spectroscopic accuracy has been attained. 
{Extrapolating 
the excitation energies along the iso-electronic sequence to estimate its value
 for the neutral atom appears to be unreliable. 
The ``divide and conquer'' strategy used in the (D)PCFI method of
 partitioning the correlation space in order to capture electron 
correlation more efficiently is confirmed to be an attractive computational 
approach.

\section*{ACKNOWLEDGMENTS}

This work was supported by the Communaut\'e fran\c{c}aise  of Belgium (Action de Recherche Concert\'ee), the Belgian National Fund for Scientific Research (FRFC/IISN Convention) and by the IUAP Belgian State Science Policy (Brix network P7/12).
 S. Verdebout is 
grateful to the ``Fonds pour la formation \`a la Recherche dans l'Industrie 
et dans l'Agriculture'' of Belgium for a PhD grant (Boursier F.R.S.-FNRS). 
P. J\"onsson, P. Rynkun, and G. Gaigalas acknowledge support from the Visby program under the Swedish Institute.


%

\end{document}